\documentclass[preprint,showpacs,preprintnumbers,amsmath,amssymb]{revtex4}

\usepackage{graphicx}% Include figure files
\usepackage{dcolumn}% Align table columns on decimal point
\usepackage{bm}% bold math

\begin{document}

\preprint{}

\title{Two-Qubit Hilbert-Schmidt Separability Functions and 
Probabilities for Full-Dimensional Even-Dyson-Index Scenarios}

\author{Paul B. Slater}% 
\email{slater@kitp.ucsb.edu}
\affiliation{%
ISBER, University of California, Santa Barbara, CA 93106\\
}%
\date{\today}% It is always \today, today,

\begin{abstract}
We extend the findings and analyses of our two recent studies 
{\it (Phys. Rev. A} {\bf{75}}, 032326 [2007] and arXiv:0704.3723) by, first, 
obtaining numerical estimates of the separability {\it function} based on the 
(Euclidean, flat) Hilbert-Schmidt (HS) 
metric for the 27-dimensional convex set of 
{\it quaternionic} two-qubit systems. 
The estimated function appears to be strongly consistent with
our previously-formulated Dyson-index ($\beta = 1, 2, 4$) 
ans{\"a}tz, dictating that 
the quaternionic ($\beta=4$) separability function should be
{\it exactly} proportional to the {\it square} of the separability function
for the 
15-dimensional convex set of 
two-qubit {\it complex} ($\beta=2$) 
systems, as well as the {\it fourth} power of the 
separability function for the 9-dimensional convex set of 
two-qubit {\it real} ($\beta=1$) systems. 
In particular, we conclude that
$\mathcal{S}^{HS}_{quat}(\mu) = (\frac{6}{71})^2 \Big((3-\mu^2) \mu\Big)^4
= (\mathcal{S}^{HS}_{complex}(\mu))^2$, $0 \leq \mu \leq 1$. Here, 
$\mu =\sqrt{\frac{\rho_{11} \rho_{44}}{\rho_{22}
\rho_{33}}}$, where $\rho$ is a $4 \times 4$ two-qubit density matrix. 
We can, thus, supplement ({\it and} fortify) 
our previous assertion that the HS separability
{\it probability} of the two-qubit {\it complex} 
states is $\frac{8}{33} \approx 0.242424$, by claiming that its quaternionic 
counterpart is $\frac{72442944}{936239725} \approx 0.0773765$. 
We also comment on and analyze the {\it odd} $\beta=1$ and 3 cases.
\end{abstract}


\maketitle
For several years now, elaborating upon an idea proposed in 
\cite{ZHSL}, we have been pursuing the problem of deriving
(hypothetically exact) formulas for the proportion of states of 
qubit-qubit and qubit-qutrit systems that are {\it separable} 
(classically-correlated) in nature 
\cite{slaterHall,slaterA,slaterC,slaterOptics,slaterJGP,slaterPRA,pbsCanosa,slaterPRA2}.
Of course, any such proportions will critically depend upon the measure 
that is placed upon 
the quantum 
systems. In particular, we have---in analogy to Bayesian
analyses, in which the {\it volume element} of the {\it Fisher information} 
metric for a parameterized family of probability distributions 
is utilized as a measure (``Jeffreys' prior'') 
\cite{kass}---principally 
employed the volume elements of the well-studied (Euclidean, flat) 
Hilbert-Schmidt (HS) and 
Bures ({\it minimal} monotone) 
metrics (as well as a number of other 
[non-minimal] {\it monotone} metrics \cite{slaterJGP}).

\.Zyczkowski and Sommers \cite{szHS,szBures} 
have, using methods of random matrix theory \cite{random}
(in particular, the Laguerre ensemble), obtained formulas,
general for all $n$, for the 
HS and Bures {\it total} volumes (and hyperareas) of $n \times n$
(real and complex)
quantum systems. Up to normalization factors, the HS total volume 
formulas were also found by Andai \cite{andai}, in a rather different 
analytical framework, using a number of 
(spherical and beta) integral identities 
and positivity (Sylvester) conditions. (He also obtained 
formulas---general for any monotone metric [including the Bures]---for the 
volume of {\it one}-qubit [$n=2$] states \cite[sec. 4]{andai}.)

Additionally, Andai did specifically study 
the HS {\it quaternionic} case. He derived the
HS total volume for $n \times n$ quaternionic systems \cite[p. 13646]{andai},
\begin{equation} \label{andaiQuatVol0} 
V_{quat}^{HS} = \frac{(2 n-2)! 
\pi^{n^2-n}}{(2 n^2-n-1)!} \Pi_{i=1}^{n-2} (2 i)!,
\end{equation}
giving us for the two-qubit ($n=4$) case of specific interest here,
the 27-dimensional volume,
\begin{equation} \label{andaiQuatVol}
\frac{\pi ^{12}}{315071454005160652800000}
\approx 2.93352 \cdot 10^{-18}.
\end{equation}
(In the analytical setting employed by \.Zyczkowski and Sommers \cite{szHS}, 
this volume would appear as 
$2^{13}$ times as large \cite[p. 13647]{andai}.)
If one then possessed a
companion volume formula for the {\it separable} 
subset, one could immediately compute 
the HS two-qubit quaternionic separability {\it probability} by taking
the ratio of the two volumes.

One analytical approach to the separable volume/probability 
question that has 
recently proved to be productive \cite{slater833}---particularly, in 
the case of the Hilbert-Schmidt (HS) metric (cf. \cite{slaterDyson})---makes 
fundamental use of a form of density matrix parameterization first
proposed by Bloore \cite{bloore}. (This methodology 
can be seen to be strongly related
to the very common and long-standing use of {\it correlation matrices} 
in statistics and its many fields of application \cite{joe,kurowicka,kurowicka2}.) 

In the Bloore 
parameterization, one simply represents an off-diagonal {\it ij}-entry
of a density matrix $\rho$, as $\rho_{ij} = \sqrt{\rho_{ii} \rho_{jj}} w_{ij}$,
where $w_{ij}$ might be real, complex or quaternionic 
\cite{asher2,adler,batle2} in nature.
The particular attraction of the Bloore scheme, in terms of the 
separability problem in which we are interested, is that one can 
(in the two-qubit case) implement
the well-known Peres-Horodecki separability (positive-partial-transpose) 
test \cite{asher,michal} 
using only the ratio $\mu =\sqrt{\frac{\rho_{11} \rho_{44}}{\rho_{22} 
\rho_{33}}}$, rather than the four (three independent) 
diagonal entries of $\rho$ individually \cite[eq. (7)]{slaterPRA2} 
\cite[eq. (5)]{slater833}.

Utilizing the Bloore parameterization, we have, accordingly, been able to reduce the problem
of computing the desired HS volumes of two-qubit separable states 
 to the computations 
of {\it one}-dimensional integrals over $\mu \in [0,\infty]$. The 
associated integrands are the 
{\it products} of
{\it two} 
functions, one a readily determined jacobian function 
$\mathcal{J}(\mu)$ (corresponding, first, 
to 
the transformation to the Bloore variables $w_{ij}$ 
and, then, to $\mu$) 
 and the other, 
the more problematical (what we have termed) 
{\it separability function} $\mathcal{S}^{HS}(\mu)$ 
\cite[eqs. (8), (9)]{slaterPRA2}.
(In the qubit-{\it qutrit} case, {\it two} 
ratios, $\mu_1$ and $\mu_2$, 
are required to express the separability conditions, 
but analytically the corresponding separability
functions also appear to be {\it univariate} in nature, being 
simply functions of $\mu_1$ or $\mu_2$, or 
the product $\mu_1 \mu_2$ \cite[sec. III]{slater833}.)

In our extensive numerical (quasi-Monte Carlo integration) 
investigation \cite{slaterPRA2} of the 9-dimensional and 15-dimensional 
convex sets of real and complex $4 \times 4$ density matrices,
 we had formulated ans{\"a}tze for the two associated separability 
functions ($\mathcal{S}^{HS}_{real}(\mu)$ and $\mathcal{S}^{HS}_{complex}(\mu)$), 
proposing that
they were 
proportional to certain (independent) 
{\it incomplete beta functions} \cite{handbook},
\begin{equation}
B_{\mu^2}(a,b) =\int_{0}^{\mu^2} \omega^{a-1} (1-\omega)^{b-1} d \omega,
\end{equation}
for particular values of $a$ and $b$.
However, in 
the subsequent study 
\cite{slater833}, 
we were led to somewhat 
modify these ans{\"a}tze, in light of multitudinous 
exact {\it lower}-dimensional results. Since
these further results clearly manifested patterns fully consistent with 
the {\it Dyson index} (``repulsion exponent'') 
pattern ($\beta =1, 2, 4$) 
of random matrix theory \cite{dyson}, we proposed
that, in the (full 9-dimensional) real case,
the separability function was proportional to a specific 
incomplete 
beta 
function ($a=\frac{1}{2},b=2$),
\begin{equation}
\mathcal{S}^{HS}_{real}(\mu) \propto B_{\mu^2}(\mu^2,\frac{1}{2},2) 
\equiv  \frac{3}{4} (3 -\mu^2) \mu
\end{equation}
and in the complex case, proportional, not to an independent function, but 
simply to the {\it square} of
$\mathcal{S}^{HS}_{real}(\mu)$.
(These proposals are strongly consistent 
\cite[Fig. 4]{slater833} with the numerical
results generated in \cite{slaterPRA2}.) This 
chain of reasoning, then, 
immediately compels one to the further proposition 
that the separability function in
the {\it quaternionic} case is exactly proportional to the {\it fourth}
power of that for the real case (and, obviously, the square of that for
the complex case). It is that specific proposition we will, first, 
seek to evaluate here.

We, thus, hope thereby to further test the validity of our 
Dyson-index ans{\"a}tz, first advanced in \cite{slater833}, 
as well as possibly develop an enlarged perspective
on the still not yet fully 
resolved problem of the HS separability probabilities 
in all three (real, complex and quaternionic) cases.
(In \cite{slater833}, we proposed, combining numerical and theoretical 
arguments, that in the real two-qubit case, the HS separability probability
is $\frac{8}{17}$, and in the complex 
two-qubit case,  $\frac{8}{33}$. 
The arguments, thusly, employed in \cite{slater833}, however, 
do not yet rise to the level of a formal
demonstration.)

Due to the ``curse of dimensionality'' \cite{bellman,kuosloan}, 
we must anticipate that for the
same number of sample ("low-discrepancy" Tezuka-Faure \cite{giray1,tezuka})
points generated in the quasi-Monte
Carlo integration 
procedure employed in \cite{slaterPRA2} and here, our numerical estimates
of the quaternionic separability function will be less precise than 
the estimates  were for
the complex, and {\it a fortiori}, real cases. 
(An interesting, sophisticated alternative
approach to computing the volume 
of {\it convex} bodies involves a variant of 
{\it simulated annealing} \cite{lovasz} (cf. \cite{dyer}), and allows one---unlike the Tezuka-Faure approach, we have so far employed---to 
establish confidence intervals for estimates.)

Our first extensive numerical 
analysis here involved the generation of sixty-four  million 24-dimensional
Tezuka-Faure points, all situated in 
the 24-dimensional unit hypercube $[0,1]^{24}$.
(The three independent 
diagonal entries of the density matrix $\rho$---being incorporated 
into the jacobian $\mathcal{J}(\mu)$---are
irrelevant at this stage of the calculations 
of $\mathcal{S}^{HS}_{quat}(\mu)$. 
The 24 [off-diagonal] Bloore variables had been
transformed so that each ranged over the unit interval [0,1]. 
The computations were done over several weeks, using compiled Mathematica 
code, on a MacMini workstation.) 

Of the sixty-four million sample points 
generated, 7,583,161, approximately 12$\%$, 
 corresponded to possible
$4 \times 4$ quaternionic density matrices---satisfying nonnegativity 
requirements. For each of these feasible points, we evaluated whether or not
the Peres-Horodecki positive-partial-transpose separability test was 
satisfied for 2,001 equally-spaced values of $\mu \in [0,1]$.

Here, we encounter another computational ``curse'', 
in addition to that already 
mentioned pertaining to the high-dimensionality of our problem, 
and also the infeasibility 
of most ($88\%$) of the sampled Tezuka-Faure points. In the standard 
manner \cite[eq. (5.1.4)]{random} 
\cite[p. 495]{adler} \cite[eq. (17)]{slaterJMP1996}  
\cite[sec. II]{JIANG}, 
making use of 
the Pauli matrices, we 
transform the $4 \times 4$ {\it quaternionic} density matrices---and their 
partial transposes---into 
$8 \times 8$ density matrices with [only] complex entries. 
Therefore, given a feasible 24-dimensional point, we have to check
for each of the 2,001 values of $\mu$, an $8 \times 8$
matrix for nonnegativity, rather than a $4 \times 4$ one, as was done in
both the real and complex two-qubit cases. In all three of these cases, 
we found that it would be incorrect
to simply assume---which would, of course, speed computations---that 
if the separability test is passed for a certain $\mu_{0}$, 
it will also be passed for all $\mu$ lying between $\mu_{0}$ and 1. 
This phenomenon reflects the intricate (quartic 
{\it both} in $\mu$ and in the Bloore variables $w_{ij}$'s, 
in the real and complex cases) nature of the
polynomial separability constraints 
\cite[eq. (7)]{slaterPRA2} \cite[eq. (5)]{slater833}.

\begin{figure}
\includegraphics{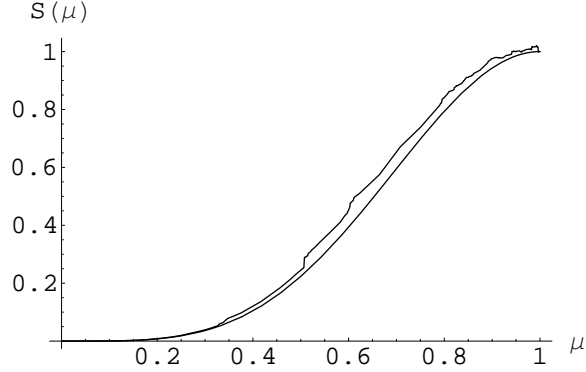}
\caption{\label{fig:quatsepfunct}Estimate---based on 64,000,000 sampled 
24-dimensional points---of the normalized form 
of the two-qubit {\it quaternionic} 
separability function, along with its (well-fitting) 
hypothetical true form, 
the {\it fourth} power of the normalized
form of $\mathcal{S}^{HS}_{real}(\mu)$, that is,
$\Big(\frac{1}{2} (3 -\mu^2) \mu\Big)^4$}
\end{figure}

In Fig.~\ref{fig:quatsepfunct} we show the estimate we, thus, were able 
to obtain
of the two-qubit quaternionic separability function 
$\mathcal{S}^{HS}_{quat}(\mu)$, in its normalized form. 
(Around $\mu=1$, one must have the evident symmetrical relation 
$\mathcal{S}^{HS}(\mu) = \mathcal{S}^{HS}(\frac{1}{\mu})$.) Accompanying our estimate 
in the plot is
the  (well-fitting) hypothetical true
form (according with our Dyson-index ans{\"a}tz 
\cite{slater833}) of the HS two-qubit separability function, 
that is, the {\it fourth} power, $\Big(\frac{1}{2} (3 -\mu^2) \mu\Big)^4$, of the 
normalized form of $\mathcal{S}^{HS}_{real}(\mu)$.

For the specific, important value of 
$\mu=1$, the ratio ($R_{1}$) of the 24-dimensional 
HS measure ($m_{sep} = R^{numer}_{1}$) 
assigned in our 
estimation procedure to separable
density matrices to the total 24-dimensional 
HS measure ($m_{tot} =R^{denom}_{1}$) 
allotted to all (separable and 
nonseparable) density matrices is $R_{1} = 0.123328$. 
The exact value of $m_{sep}$ is, of course, to begin here, unknown, being
a principal desideratum of our investigation. On the other hand, 
we can directly deduce that 
$m_{tot} = R_{1}^{denom} = 
\frac{\pi ^{12}}{7776000} \approx 0.118862$---our sample 
estimate  being 0.115845---by dividing the two-qubit HS quaternionic 
27-dimensional volume 
(\ref{andaiQuatVol}) by
\begin{equation} \label{rationalfraction}
R_{2}^{denom} =2 \int_{0}^{1} \mathcal{J}_{quat}(\mu) d \mu =
\frac{\Gamma \left(\frac{3 \beta }{2}+1\right)^4}{\Gamma
   (6 \beta +4)} =
\frac{1}{40518448303132800} \approx 2.46801 \cdot 10^{-17}, 
\hspace{.1in} \beta  = 4.
\end{equation}

\begin{figure}[!tbp]
\includegraphics{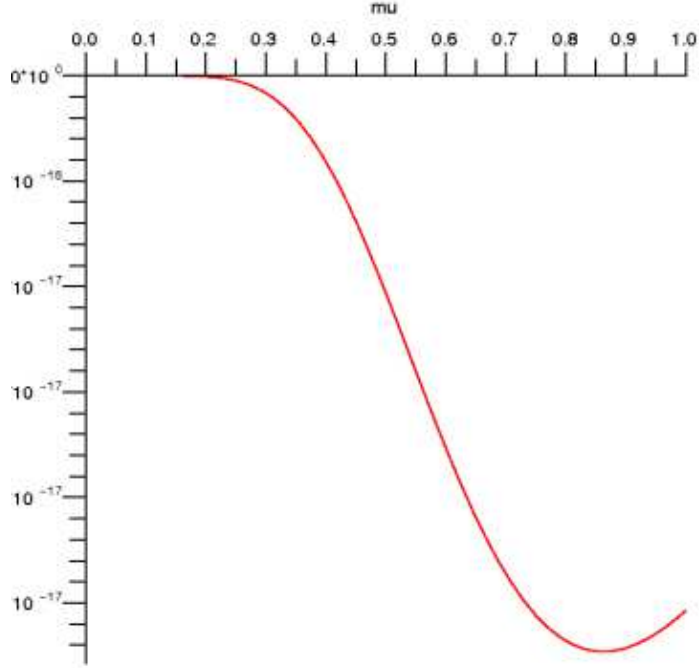}
\caption{\label{fig:quatjacobian}The univariate quaternionic jacobian function 
$\mathcal{J}_{quat}(\mu)$}
\end{figure}

Here, $\mathcal{J}_{quat}(\mu)$ is the quaternionic jacobian function
 (Fig.~\ref{fig:quatjacobian}),
obtained by transforming the
quaternionic Bloore
jacobian $\Big(\rho_{11} \rho_{22} \rho_{33} 
(1-\rho_{11} -\rho_{22} -\rho_{33})\Big)^\frac{3 \beta}{2}$, $\beta=4$, 
to the $\mu$ variable by replacing, say $\rho_{33}$ by $\mu$,
and integrating out $\rho_{11}$ and $\rho_{22}$.
(We had presented plots of $\mathcal{J}_{real}(\mu)$ 
and $\mathcal{J}_{complex}(\mu)$ 
in \cite[Figs. 1, 2]{slaterPRA2}, and observed 
apparently highly oscillatory behavior in both functions in 
the vicinity of $\mu=1$. 
However, a referee of \cite{slater833} informed us that this was simply
an artifact of using standard machine precision, and that with 
sufficiently enhanced
precision, the oscillations could be seen to be, in fact, illusory.)
We can obtain an 
estimate of the two-qubit 
quaternionic {\it separability} 
probability $P^{HS}_{sep/quat}$ by multiplying the ratio
$R_{1}$ by a second ratio $R_{2}$.
The denominator of $R_{2}$ has already been given
(\ref{rationalfraction}).
The {\it numerator} of $R_{2}$ is the specific value
\begin{equation} \label{R2numerator}
R_{2}^{numer} = 2 \int_{0}^{1} \mathcal{J}_{quat}(\mu) 
\Big(\frac{1}{2} (3 -\mu^2) \mu\Big)^4 d \mu = \frac{5989}{358347086242825680000} 
\approx 1.67128 \cdot 10^{-17},
\end{equation}
where, to obtain the integrand, 
we have multiplied (in line with our basic 
[Bloore-parameterization] approach to the separability 
probability question) 
the quaternionic jacobian function by the 
(normalized) putative 
form of the two-qubit quaternionic separability function. 
(Note the use of the $\beta=4$ exponent.)

The counterpart of $R^{numer}_{2}$ in the 9-dimensional real case is  
$\frac{1}{151200}$ and in the 15-dimensional complex case, 
$\frac{71}{99891792000}$. We now note that 
\begin{equation}
99891792000 = \left(
\begin{array}{c}
 \text{11} \\
 \text{2}
\end{array}
\right) \frac{\Gamma{(16)}}{\Gamma{(7)}}
\end{equation}
is the coefficient 
of $\mu^2$ in $11 !  
L_{11}^{4}(\mu)$ and $\frac{151200}{2}= 75600$ plays the exact same role 
in $6! L_{6}^{4}(\mu)$, where $L_{m}^{4}(\mu)$ is a generalized ($a=4$)
Laguerre polynomial  (see sequences A062260 and A062140  
in the {\it The On-Line Encylopaedia of Integer
Sequences}). (Also, as regards the denominator of (\ref{R2numerator}), 
$\frac{358347086242825680000}{3587352665} = 99891792000$.)
\.Zyczkowski and Sommers had 
made use of the Laguerre ensemble in deriving
the HS and Bures volumes and hyperareas of $n$-level quantum systems 
\cite{szHS,szBures}. Generalized (associated/Sonine) Laguerre 
polynomials (``Laguerre functions'') have been employed in, 
in another important quantum-information context, in 
proofs of Page's conjecture on the average entropy of a subsystem 
\cite{sanchez,sen}.)

We, thus, have, for our two-qubit quaternionic case, that
\begin{equation} \label{ratio2}
R_{2} = \frac{R_{2}^{numer}}{R_{2}^{denom}} = 
\frac{125769}{185725} \approx 0.677179.
\end{equation}
(The {\it real} counterpart of $R_{2}$ is
$\frac{1024}{135 \pi ^2} \approx 0.76854$, and the 
{\it complex} one, $\frac{71}{99} \approx 0.717172$. 
Additionally, we computed that the 
corresponding ``truncated'' quaternionic \cite{pfaff}
ratio---when {\it one} of the four quaternionic parameters is set to 
zero, that is the Dyson-index 
case $\beta=3$--- is $\frac{726923214848}{106376244975 
\pi ^2} \approx 0.692379$. Thus, we see that these four 
important ratios monotonically 
decrease as $\beta$ increases, and also, significantly, that the two 
ratios for odd values of $\beta$ 
differ qualitatively---both having $\pi^2$ in their denominators---from 
those two for even $\beta$.)

Our quasi-Monte Carlo (preliminary)
estimate of the two-qubit quaternionic separability 
{\it probability} is, then, 
\begin{equation} \label{r1r2}
P_{sep/quat}^{HS} \approx R_{1} R_{2} =0.0813594.
\end{equation}
Multiplying the total volume of the 27-dimensional convex set of 
two-qubit quaternionic states, given in the framework 
of Andai \cite{andai} by (\ref{andaiQuatVol}), by this result 
(\ref{r1r2}), we 
obtain the two-qubit quaternionic separable volume 
estimate $V^{HS}_{sep/quat} \approx  2.38775 \cdot 10^{-19}$. 

Our 24-dimensional quasi-Monte Carlo integration procedure 
leads to a derived estimate of (the total 27-dimensional volume) 
$V^{HS}_{quat}$, that was
somewhat smaller, $2.85906
\cdot 10^{-18}$, than the $2.93352 \cdot 10^{-18}$ 
given by (\ref{andaiQuatVol}). 
Although rather satisfying, this 
was sufficiently imprecise to discourage us
from attempting to ``guestimate'' the 
(all-important) constant ($R_{1}$) by which to multiply
the putative normalized form, $(\frac{1}{2} (3-\mu^2) \mu)^4$, 
of the quaternionic separability function in (\ref{R2numerator}) 
in order to yield
the true separable volume.
In our previous study \cite[sec. IX.A]{slater833}, we presented certain plausibility
arguments to the effect that
the corresponding 
constant in the 9-dimensional real case might be 
$\frac{135 \pi^2}{2176} = (\frac{20 \pi^4}{17})/(\frac{512 \pi^2}{27})$, and 
$\frac{24}{71} =(\frac{256 \pi^6}{639})/(\frac{32 \pi^6}{27})$ in the 15-dimensional complex case.
(This leads---multiplying by the corresponding $R_{2}$'s, 
$\frac{1024}{135 \pi^2}$ and $\frac{71}{99}$---to 
separability probabilities of $\frac{8}{17}$ and 
$\frac{8}{33}$, respectively.)

In light of such imprecision, we undertook a supplementary 
analysis, in which, instead of examining each feasible 24-dimensional point
for 2,001 possible values of $\mu$, with respect to separability or not, 
we simply used $\mu=1$. This, of course,
allows us to significantly increase the number of 24-dimensional
Tezuka-Faure points generated from the 64,000,000 so far employed. 

We, thusly, generated 1,360,000,000 points, finding 
that we obtained a remarkably good fit to the important ratio
$R_{1}$ of the 24-dimensional measure, at $\mu=1$, assigned to the separable
two-qubit quaternionic density matrices to the measure (known to be
$\frac{\pi^{12}}{7776000}$) by setting $R_{1} =(\frac{24}{71})^2 
\approx 0.114263$ (our sample estimate of this quantity 
being 0.114262).  This is 
{\it exactly} the square of the corresponding ratio $\frac{24}{71}$ we had
conjectured (based on extensive numerical and theoretical evidence) for 
the full (15-dimensional) complex two-qubit case
in \cite{slater833}.

Under this hypothesis on $R_{1}$, we have the ensuing 
string of relationships
\begin{equation}
\mathcal{S}^{HS}_{quat}(\mu) = \Big(\frac{24}{71})^2 (\frac{1}{2} (3 -\mu^2) \mu\Big)^4 
=  \Big(\frac{6}{71}\Big)^2 \Big( (3 -\mu^2) \mu\Big)^4
=\Big( \mathcal{S}^{HS}_{complex}(\mu) \Big)^2,
\end{equation}
with (as already advanced in \cite{slater833})
\begin{equation}
\mathcal{S}^{HS}_{complex}(\mu) = 
\frac{24}{71} \Big(\frac{1}{2} 
(3-\mu^2) \mu\Big)^2= \frac{6}{71} \Big((3-\mu^2) \mu\Big)^2.
\end{equation}
Then, using our knowledge of the complementary ratio $R_{2}$, given in 
(\ref{ratio2}), we obtain
\begin{equation}
P^{HS}_{sep/quat} = R_{1} R_{2} = \frac{72442944}{936239725} 
\approx 0.0773765,
\end{equation}
as well as---in the framework of Andai \cite{andai}---that
\begin{equation}
 V^{HS}_{sep/quat} =\frac{5989 \pi ^{12}}{24386773433626137413880000000} 
\approx 2.26986 \cdot 10^{-19}.
\end{equation}

For possible further insight into the HS two-qubit separability 
probability question,
we undertook a parallel quasi-Monte Carlo (Tezuka-Faure) integration 
(setting $\mu=1$) for
the truncated quaternionic case ($\beta=3$), in which one of the four
quaternionic parameters is set to zero. Although there was 
no corresponding formula
for the HS total volume for this scenario given in \cite{andai}, 
upon request, A. Andai 
kindly derived the result
\begin{equation} \label{puzzling}
V^{HS}_{trunc} = \frac{\pi ^{10}}{384458588946432000} 
\approx 2.43584 \cdot 10^{-13}.
\end{equation}
(In fact, Andai was able to derive {\it one} simple 
overall comprehensive formula---which we leave for him to publish---yielding
the total HS volumes for all $n \times n$ systems and Dyson indices $\beta$.)
Let us, further, note that Andai obtains 
the result (\ref{puzzling}) as the product of three factors, 
$V^{HS}_{trunc} =\pi_1 \pi_2 \pi_3$, where
\begin{equation}
\pi_1 = \frac{128 \pi ^8}{105}; \hspace{.1in} \pi_2 =\frac{128}{893025}; 
\hspace{.1in} 
\pi_3=\frac{189 \pi ^2}{12696335643836416}.
\end{equation}

Now, we will simply {\it assume}---in line with our basic Dyson-index 
ans{\"a}tz, substantially supported in 
\cite{slater833} and above---that 
the corresponding separability function is of the form
\begin{equation}
\mathcal{S}^{HS}_{trunc}(\mu) \propto 
( (3 -\mu^2) \mu)^\beta, \hspace{.15in} \beta=3.
\end{equation}
(Of course, one should ideally test this 
specific application of the ans{\"a}tz too, perhaps in the manner we have 
examined the $\beta =4$ instance above [Fig.~\ref{fig:quatsepfunct}].)

We were somewhat perplexed, however, by the results of our quasi-Monte Carlo
integration procedure, conducted in the 18-dimensional space of 
off-diagonal entries of the truncated quaterionic density matrix 
$\rho$. Though, we 
anticipated (from our previous 
extensive numerical experience here and elsewhere) that 
our estimate of the associated 18-dimensional volume would be, at least, 
within a few percentage points  of $\pi_1 \pi_2 = 
\frac{16384 \pi ^8}{93767625} \approx 
1.65793$, the estimate was, in fact, close to 0.967 (1, thus, falling within the possible margin of error).
Assuming the correctness of the analysis of Andai, which we have no other 
reason to doubt, the only possible explanations seemed to be that we had
committed some programming error (which 
we were unable to discern) or that we had
some conceptual misunderstanding regarding the analysis of truncated
quaternions. (Let us note that we do convert the $4 \times 4$ density matrix
to $8 \times 8$ [complex] form 
\cite[p. 495]{adler} \cite[eq. (17)]{slaterJMP1996}
\cite[sec. II]{JIANG}, while it appears that Andai does not 
directly employ such a transformation in his derivations.)

In any case, we did 
devote considerable computing time to the $\beta=3$ problem 
(generating 1,180,000,000 18-dimensional Tezuka-Faure 
points), with the hope being
that if we were in some way in error, the error would be 
an {\it unbiased} 
one, and 
that the all-important {\it ratio} of volumes would be unaffected.

Proceeding thusly, our best estimate 
({\it not} making use of the Andai result (\ref{puzzling}) for the present) 
of the HS separability probability 
was 0.193006. One interesting possible candidate exact value
is, then, $\frac{128}{633} = \frac{2^7}{3 \cdot 13 \cdot 17} 
\approx 0.193062$. (Note the presence of 128 in the numerators, also, 
of both factors 
$\pi_1$ and $\pi_2$.)
This would give us a 
counterpart [$\beta=3$] value for the ratio $R_{2}$ of 
$\frac{160446825 \pi ^2}{5679087616} \approx 0.278838$. 
In \cite{slater833}, we had asserted that, in the other 
odd $\beta=1$ case, the counterpart of $R_{2}$ was
$\frac{135 \pi ^2}{2176} \approx 0.612315$. (Multiplying this by 
$\frac{1024}{135 \pi^2}$ gave us the 
conjectured HS {\it real} two-qubit separability probability 
of $\frac{8}{17}$.)

So, let us say in conclusion, that although we believe we have successfully
resolved---though still far from having formal proofs---the 
two-qubit Hilbert-Schmidt 
separability probability question for the $\beta =2$ and 4 
(complex and quaternionic) cases, the odd ($\beta =1, 3$) cases, in 
particular $\beta =3$, appear still to be somewhat 
more problematical.

\begin{acknowledgments}
I would like to express gratitude to the Kavli Institute for Theoretical
Physics (KITP)
for computational support in this research.

\end{acknowledgments}

\bibliography{Quaternionic5}% Produces the bibliography via BibTeX.

\end{document}